\documentclass[runningheads]{llncs}
\usepackage{hyperref}
\usepackage[T1]{fontenc}
\usepackage{graphicx}
\usepackage{cite}
\usepackage{color}
\usepackage{amsmath}
\usepackage{amssymb}
\usepackage{sidecap}

\begin{document}

\title{Social Fragmentation Transitions in Large-Scale Parameter Sweep Simulations of Adaptive Social Networks\thanks{This work was supported in part by JSPS KAKENHI Grant Number 19K21571.}}
\titlerunning{Social Fragmentation Transitions in Adaptive Social Networks}
%
\author{Hiroki Sayama\inst{1,2}\orcidID{0000-0002-2670-5864}}
\authorrunning{Hiroki Sayama}
%

\institute{Center for Collective Dynamics of Complex Systems, Binghamton University,\\State University of New York, Binghamton, NY 13902-6000, USA \and
Waseda Innovation Lab, Waseda University, Shinjuku, Tokyo 169-8050, Japan
\email{sayama@binghamton.edu}}
\maketitle              
\begin{abstract}
Social fragmentation transition is a transition of social states between many disconnected communities with distinct opinions and a well-connected single network with homogeneous opinions. This is a timely research topic with high relevance to various current societal issues. We had previously studied this problem using numerical simulations of adaptive social network models and found that two individual behavioral traits, homophily and attention to novelty, had the most statistically significant impact on the outcomes of social network evolution. However, our previous study was limited in terms of the range of parameter values examined, and possible interactions between multiple behavioral traits were largely ignored. In this study, we conducted a substantially larger-scale parameter sweep numerical experiment of the same model with expanded parameter ranges by an order of magnitude in each parameter dimension, resulting in a total of 116,640 simulation runs. To capture nontrivial interactions among behavioral parameters, we modeled and visualized the dependence of outcome measures on the model parameters using artificial neural networks. Results show that, while the competition between homophily and attention to novelty is still the primary determinant of social fragmentation, another transition plane emerges when individuals have strong social conformity behavior, which was not previously known. This implies that social fragmentation transition can also occur in the homophily-social conformity trade-off, the two behavioral traits that have very similar microscopic individual-level effects but produce very different macroscopic collective-level outcomes, illustrating the nontrivial macroscopic dynamics of complex collective systems.
\keywords{adaptive social networks  \and
social fragmentation \and 
large-scale numerical simulations \and
homophily \and
attention to novelty \and
social conformity.}
\end{abstract}
\section{Introduction}

The study of temporal evolution of social structure is one of the significant application domains of complex collective systems research. In particular, social fragmentation transition, i.e., transition of social states between many disconnected communities with distinct opinions and a well-connected single network with homogeneous opinions, is a timely research topic with high relevance to various current societal issues \cite{holme2006,zanette2006,kozma2008,bohme2011,sayama2020b,blex2020,levin2021,sasahara2021}. We had previously studied this problem using numerical simulations of adaptive social network models \cite{sayama2020} and found that two individual behavioral traits, homophily (i.e., tendency to strengthen connections to similar individuals and weaken those to dissimilar ones) \cite{mcpherson2001,kossinets2009,bakshy2015} and attention to novelty (i.e., tendency to strengthen connections to individuals whose opinions stand out compared to others), had the most statistically significant impact on the outcomes of social network evolution \cite{sayama2020}. Specifically, when homophily was strong, the social network evolved into fragmented states of many disconnected clusters with diverse opinions, but when attention to novelty was strong, the social network evolved to well-connected yet informationally homogeneous states. However, the previous study was rather limited in terms of the range of parameter values examined, and possible interactions between multiple behavioral traits were largely ignored, especially about the other behavioral trait, social conformity (i.e., how strongly individuals assimilate themselves to social neighbors).

In this study, we examined a broader spectrum of social network dynamics through a larger-scale parameter sweep experiment of the same model with expanded parameter ranges by an order of magnitude in each parameter dimension, resulting in a total of 116,640 simulation runs. To capture nontrivial interactions among behavioral parameters, we modeled and visualized the dependence of outcome measures on the model parameters using artificial neural networks. Results show that, while the competition between homophily and attention to novelty is still the primary determinant of social fragmentation when social conformity behavior of individuals is weak, another transition plane emerges at an intermediate homophily level when individuals have strong social conformity behavior, which was not previously known. 

In what follows, we describe the model of adaptive social networks and the settings and results of the large-scale parameter sweep numerical experiments. We further discuss implications of the results for social evolution and potential future research directions.

\section{Model}

Our original model \cite{sayama2020} describes distributed opinion dynamics on an adaptive
social network made of $n$ nodes. Adaptive networks \cite{gross2009,sayama2013camwa} are a class of dynamical network models in which node states and edge connectivities co-evolve in adaptation to each other. In our model, node $i$ has its own opinion state $x_i \in \mathbb{R}$. Nodes are connected through 
weighted directed edges that represent the information flow from source to target nodes. The edge weight is denoted as $w_{ij} \in
\mathbb{R}_{\geq 0}$, where $i$ is the target node and $j$ is the source node. 

The adaptive network dynamics, i.e., the co-evolution of node states and edge weights, are governed by the following differential equations:
\begin{align}
\frac{d x_i}{d t} &= c \left( \langle x \rangle_i - x_i \right) + \epsilon \label{dxdt}\\
\frac{d w_{ij}}{d t} &= h F_h(x_i, x_j) + a F_a(\langle x \rangle_i, x_j) \label{dwdt}\\
\langle x \rangle_i &= \frac{\sum_j w_{ij} x_j}{\sum_j w_{ij}} \label{localav}
\end{align}
Here, $\langle x \rangle_i$ (Eq.~(\ref{localav})) represents the weighted local average of neighbors' opinions (i.e., social norm) perceived by node $i$.
Parameter $c$ and noise term $\epsilon$ in Eq.~(\ref{dxdt}) represent the strength of social conformity and stochastic fluctuation of node states, respectively.
Parameters $h$ and $a$ in Eq.~(\ref{dwdt}) represent the strength of homophily and attention to novelty, respectively.
$F_h$ and $F_a$ in Eq.~(\ref{dwdt}) are functions that describe the increase/decrease of edge weights because of homophily and attention to novelty, respectively. $F_h$ and $F_a$ can be any functions that monotonically decrease (for $F_h$) or increase (for $F_a$) as the distance between the two arguments increase. In this study, we used the following simple functions for $F_h$ and $F_a$:
\begin{align}
F_h(x_i, x_j) &= \theta_h - |x_i - x_j| \\
F_a(\langle x \rangle_i, x_j) &= |\langle x \rangle_i - x_j| - \theta_a
\end{align}
Here $\theta_h$ and $\theta_a$ are the default values of $F_h$ and $F_a$, respectively, when the two given arguments are equal. These functions describe that the edge from node $j$ to node $i$
tends to become strengthened when $j$'s state is similar to $i$'s (i.e., homophily) and distant from the local average (i.e.,
attention to novelty), or weakened otherwise. We restricted $w_{ij}$
to be always nonnegative, and any negative values resulted from numerical simulation of Eq.~(\ref{dwdt}) would be rounded up to zero.

\begin{SCfigure}[][t]
\centering
\includegraphics[width=3in]{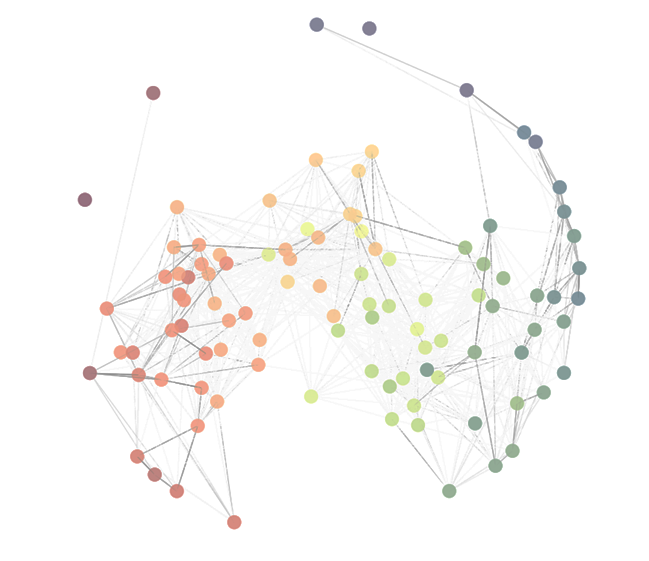}
\caption{A snapshot of the proposed adaptive social network model visualized in the middle of a simulation. Colors of nodes represent their states (opinions) using Matplotlib's ``spectral'' color map, while the shade of edges represent their weights (connection strengths).}
\label{screenshot}
\end{SCfigure}

Simulating this adaptive social network model from a random initial condition produces a sequence of social network configurations in which node states (opinions) spread through social ties and edge weights (connection strengths) also change due to node states (an example is shown in Fig.~\ref{screenshot}). We implemented the numerical simulator of the model in Python 3.7
with NetworkX \cite{hagberg2008} and PyCX \cite{sayama2013}\footnote{The simulator code is available from the
  author upon request.}.
  
\begin{figure}[tp]
\centering
\includegraphics[width=\columnwidth]{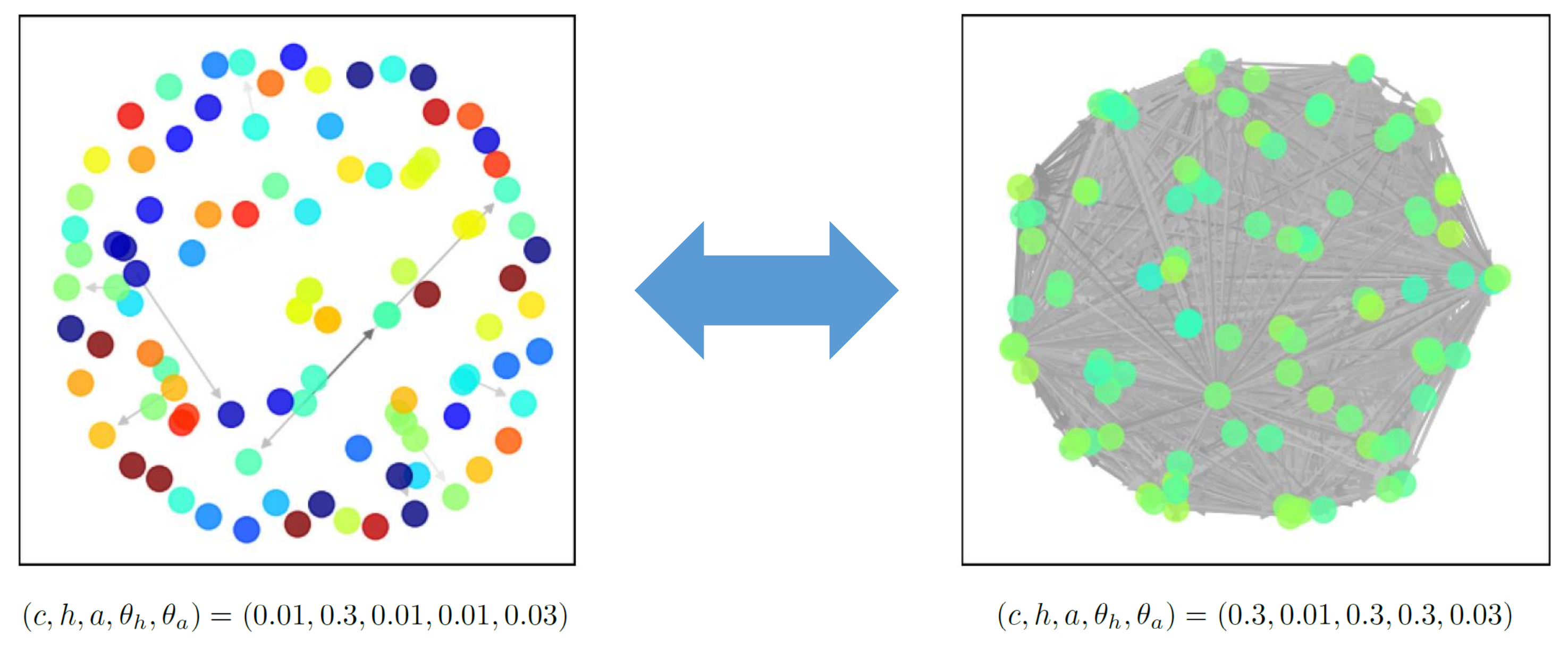}
\caption{Examples of final states of the adaptive social network simulation with $n=100$ that demonstrate social fragmentation transition. Left: Fragmented state with large $h$ and small $a$. Right: Homogenized state with small $h$ and large $a$. Specific parameter settings are shown beneath each panel. Visualizations were done in the same way as in Fig.~\ref{screenshot}.}
\label{transition}
\end{figure}

This model is known to exhibit social fragmentation transition, i.e., transition between fragmented and homogenized social network states, as the individuals' behavioral parameters are varied (Fig.~\ref{transition}) \cite{holme2006,zanette2006,kozma2008,bohme2011}. Our previous study \cite{sayama2020} showed that, when homophily ($h$) is stronger or attention to novelty ($a$) is weaker, the social network is more inclined to become fragmented into many disconnected small clusters with various opinion states (Fig.~\ref{transition}, left), and in the opposite settings social homogenization is more likely to occur (Fig.~\ref{transition}, right). Meanwhile, the potential effect of social conformity ($c$) was unclear in the previous analysis, which is the main focus of the present study.

\section{Experiments}

\subsection{Settings}

We conducted numerical simulations of the above adaptive social network model to systematically investigate the effects of individual behavioral parameters ($c$, $h$, $a$, $\theta_h$, and $\theta_a$) on the course of social network evolution. The parameter values used are as follows:
\begin{itemize}
\item Network size: $n \in \{30, 100, 300\}$\footnote{Simulations with a larger network size ($n=1,000$) were also conducted in our earlier work \cite{sayama2020} and we confirmed that their results did not differ much from those with $n=300$.}
\item Behavioral parameters: $c, \; h, \; a, \; \theta_h, \; \theta_a \in \{0.003, 0.01, 0.03, 0.1, 0.3, 1.0\}$
\end{itemize}
The above range of values for behavioral parameters was an order-of-magnitude larger in each dimension than what was examined before \cite{sayama2020}. Each parameter value combination was simulated 5 times with independently generated random initial conditions. This resulted in a total of $3 \times 6^5 \times 5 = 116,640$ simulation runs, taking a substantial amount of computational time and resource. Simulations were thus conducted in parallel on four designated PCs for over a few months.

In each simulation, the initial configuration of the network was randomly generated so that every pair of nodes were connected by two directed edges (in both directions) with a randomly generated weight sampled from a standard uniform distribution ($w_{ij} \in [0, 1]$) in each direction\footnote{We did not use more realistic social network structures like those with long-tailed degree distributions or modular community structures. This is because, in order to understand social self-organization, those structures should arise as an {\em outcome} of dynamical interactions among agents rather than used as the initial condition given {\em a priori}.} and each node had a random node state sampled from the normal distribution ${\cal N}(0, 1)$. Equations (\ref{dxdt}) and (\ref{dwdt}) were numerically simulated using a simple Euler forward method with
time step size $\Delta t = 0.1$ for $t \in [0, 100]$. The stochastic behavior of node states represented by $\epsilon$ in Eq.~(\ref{dxdt}) was simulated by adding a random number
sampled from ${\cal N}(0, 0.1^2)$ to $x_i$ at each discrete time step $\Delta t$.

\subsection{Outcome measures}

At the end of each simulation run ($t=100$), we converted the final network configuration into an undirected network by replacing the two directed edges between each pair of nodes with a single undirected edge whose weight was the average of the original two edges' weights. Then the Louvain modularity maximization method \cite{blondel2008fast} was applied to the undirected network to detect community structure in the final network configuration. Within each detected community, we calculated the average node state (called
``average community state'' hereafter). Using the results of these steps, we calculated the following five 
network metrics as final outcome measures:
\begin{enumerate}
\item Average edge weight (= arithmetic average of all the edge
  weights in the network)
\item Number of communities
\item Modularity of the community structure
\item Range of average community states (= difference between largest
  and smallest average community states)
\item Standard deviation of average community states
\end{enumerate}
These outcome measures were averaged over five independent simulation runs for each combinations of parameter values. The first three outcome measures capture the structural properties of the social network, while the last two capture the opinion diversity in the social network. When the social network is fragmented, the average edge weight takes a small value, while all the other measurements takes large values. The opposite pattern is realized when the social network is homogenized. This allows us to easily detect which state the adaptive social network evolved into in quantitative ways.

\section{Results}

In order to capture and visualize the effects of the five behavioral parameters on the five outcome measures (including possible nonlinear interactions among those behavioral parameters), we modeled the parameter-outcome mapping using artificial neural networks with Wolfram Research Mathematica 12's artificial neural network predictor \cite{mathematica}. Natural logarithms of the five behavioral parameter values were used as five-dimensional input vectors, and the five outcome measures obtained from simulation results under those parameter settings were used as five-dimensional output vectors. The combinations of these input and output vectors were gathered for the whole simulation runs for each network size ($n \in \{30, 100, 300\}$) and used as the data set to train an artificial neural network model for specific $n$. The performance goal of training was set to maximizing the accuracy of outcome prediction \cite{mathematica}.

\begin{figure}[tp]
\includegraphics[width=\columnwidth]{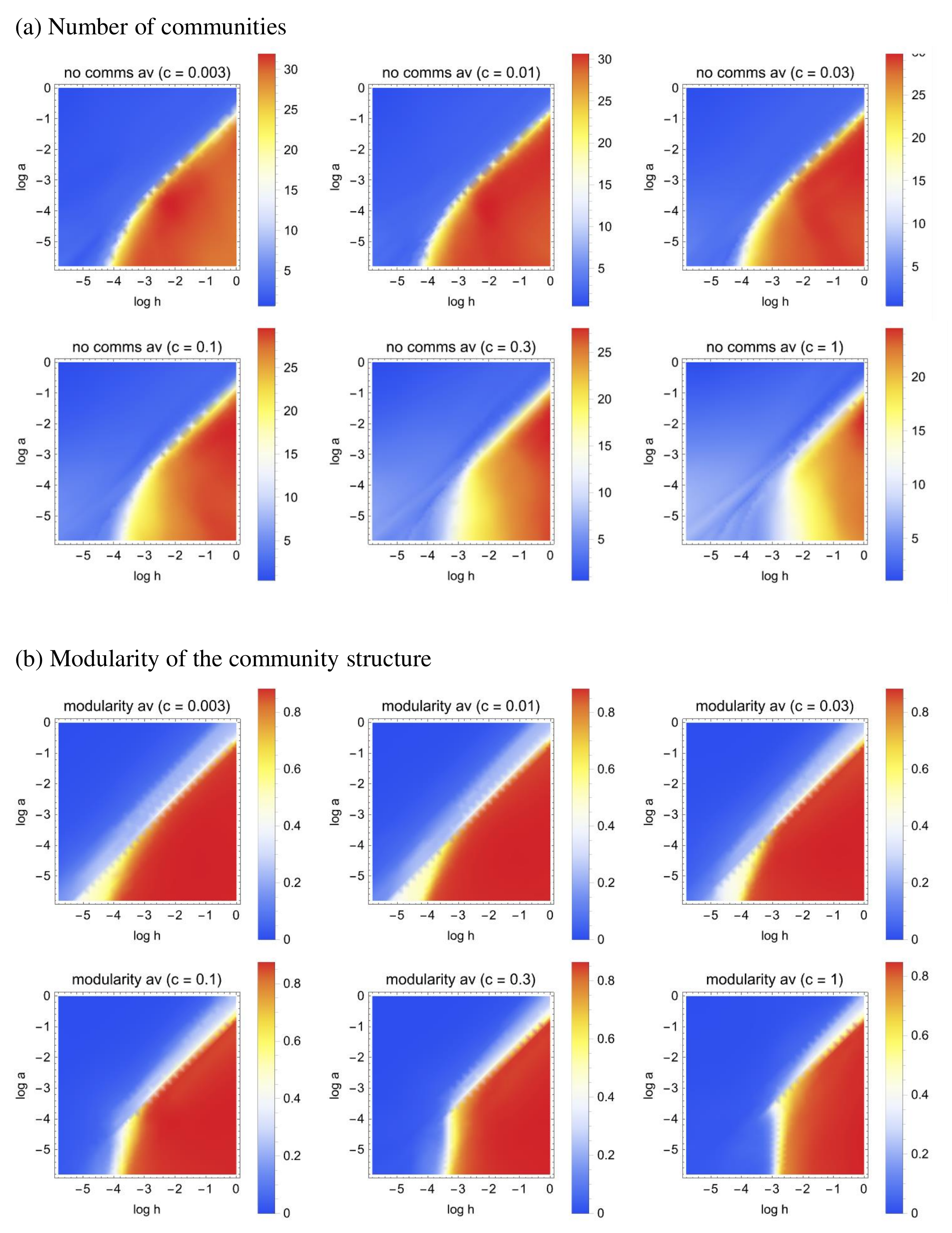}
\caption{Phase diagrams of adaptive social network evolution in terms of network structure outcome measures. Each plot shows outcome dependence on homophily ($h$, horizontal axis), attention to novelty ($a$, vertical axis) and conformity ($c$, varied from top-left to bottom-right) modeled using artificial neural networks. (a) How the number of communities depends on $h$, $a$ and $c$. (b) How the modularity of the community structure depends on $h$, $a$ and $c$. Red and blue regions correspond to fragmented and homogenized network states, respectively. $n = 300$, $\theta_h = 0.1$, and $\theta_a = 0.1$. Similar patterns were observed for other outcome measures and network sizes.}
\label{phase-spaces-structure}
\end{figure}

\begin{figure}[tp]
\includegraphics[width=\columnwidth]{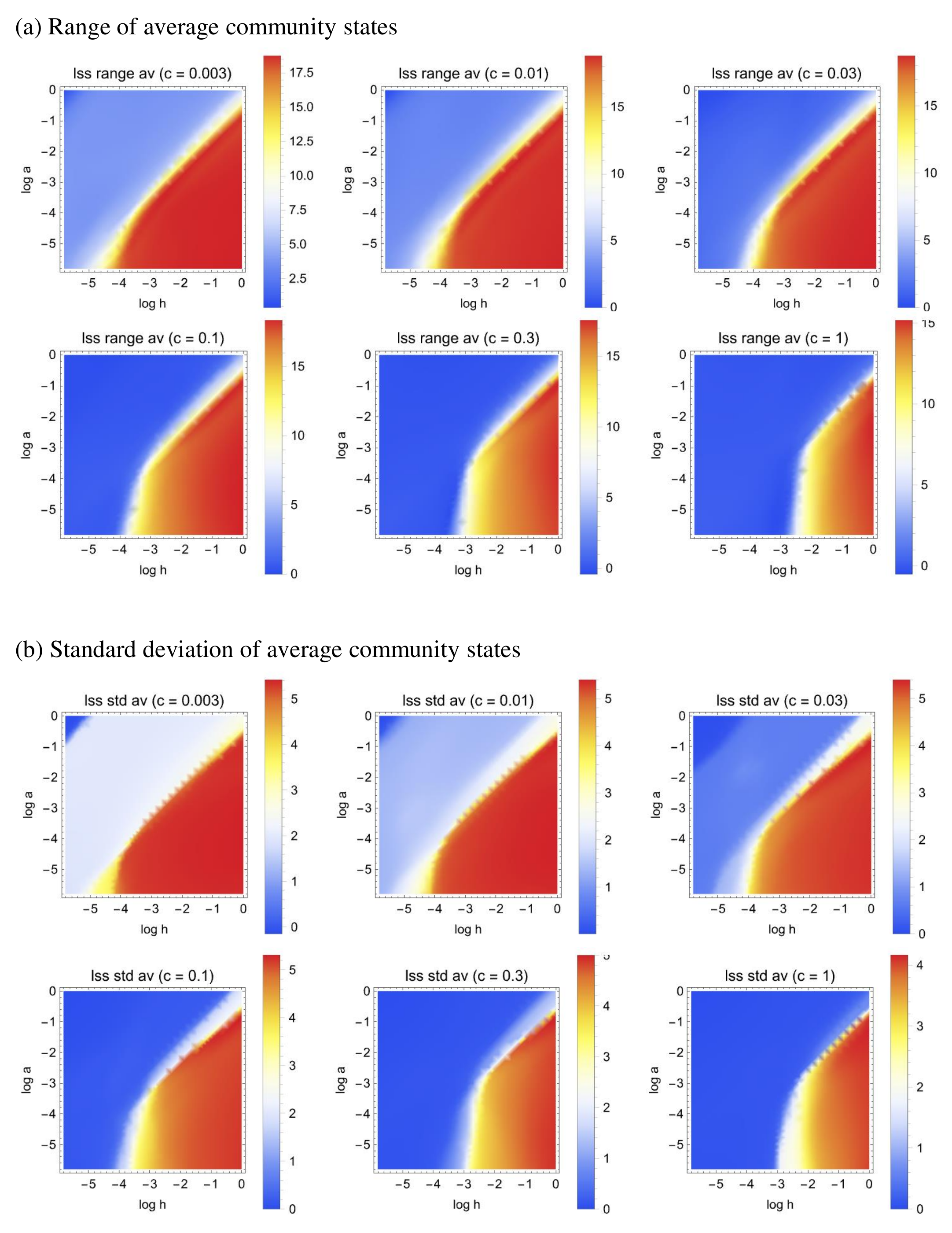}
\caption{Phase diagrams of adaptive social network evolution in terms of node state outcome measures. Each plot shows outcome dependence on homophily ($h$, horizontal axis), attention to novelty ($a$, vertical axis) and conformity ($c$, varied from top-left to bottom-right) modeled using artificial neural networks. (a) How the range of average community states depends on $h$, $a$ and $c$. (b) How the standard deviation of average community states depends on $h$, $a$ and $c$. Red and blue regions correspond to fragmented and homogenized network states, respectively. $n = 300$, $\theta_h = 0.1$, and $\theta_a = 0.1$. Similar patterns were observed for other outcome measures and network sizes.}
\label{phase-spaces-state}
\end{figure}

Illustrative results with $n=300$ are shown as heat maps of each outcome measure in Fig.~\ref{phase-spaces-structure} (for final network structure) and Fig.~\ref{phase-spaces-state} (for final node states). 
The competition between homophily ($h$) and attention to novelty ($a$) is still observed as the primary determinant of social fragmentation in a low-conformity ($c$) regime (top rows of all panels in Figs.~\ref{phase-spaces-structure} and \ref{phase-spaces-state}), seen as the diagonal transition plane in the plots. However, another vertical transition plane emerges at an intermediate homophily level in a high-conformity regime (bottom rows of all panels in Figs.~\ref{phase-spaces-structure} and \ref{phase-spaces-state}), which was not previously known. Similar patterns were observed for other outcome measures and network sizes. This new result shows that, when individuals' social conformity ($c$) is sufficiently strong, homogenization of the social network can occur even without attention to novelty. This implies that social conformity and attention to novelty, while very different in their intentions and actions at microscopic individual levels, have similar effects of promoting connections among individuals in an adaptive social network.

The result shown above also reveals a previously unrecognized competition between social conformity ($c$) and homophily ($h$) when attention to novelty is weak (i.e., low-$a$ regions; near the bottom edge of each heat map in Figs.~\ref{phase-spaces-structure} and \ref{phase-spaces-state}). Namely, when $c$ is low social fragmentation dominates, but when $c$ is high social homogenization becomes possible for smaller values of $h$. This is quite intriguing because these two behaviors (social conformity and homophily) have very similar effects at an individual level (i.e., they both make ego and alter similar to each other). In fact, their differences are often very vague and undetectable in empirical social network studies \cite{shalizi2011}. Meanwhile, these two behaviors are mechanistically distinct, because social conformity is about node dynamics while homophily is about edge dynamics. This finding, that their competitive balance may lead to very different societal outcomes down the road, offers a lot of implications for how we should consider our social interactions and behaviors in this highly interconnected world.

\section{Conclusions}

In this study, we conducted large-scale parameter sweep simulations of our adaptive social network model to investigate the transition points between fragmentation and homogenization of social networks in a multidimensional behavioral parameter space. Artificial neural network-based modeling and visualization of the parameter-outcome mapping revealed a new transition plane for strong social conformity ($c$) and weak attention to novelty ($a$) regimes, which was previously unrecognized. The overall multidimensional phase space structure shows a nonlinear interaction among the three key behavioral mechanisms (social conformity, homophily, and attention to novelty). Within the range of parameter values tested so far, it appears that social homogenization (blue regions in Figs.~\ref{phase-spaces-structure} and \ref{phase-spaces-state}) occupied a greater volume in the log-scale parameter space than social fragmentation did.

This study presents a concrete example of complex collective systems research to study {\em Artificial Society}, i.e., study of hypothetical models of {\em society-as-it-could-be}. Such theoretical/mathematical/computational exploration of social systems can play valuable roles complementary to more empirical social science research, in the same spirit of Artificial Life research \cite{langton1992} that complements traditional biology. Computational examination of hypothetical scenarios, such as changing individual behaviors in our model, allows for exploration of various possible forms of our society and may lead to a discovery of novel possible social states which would not be realized just by analyzing empirical data obtained from real society \cite{sayama2020b}. Such exploratory endeavor is becoming increasingly important and relevant in today's highly automated, interconnected society, as our daily interactions are moving away from traditional, ``natural'' forms and becoming more and more mediated by artificially designed, ``engineered'' communication platforms. This has become even more manifested because of the recent COVID-19 pandemic (think about Zoom, YouTube, Slack, and other social media/collaboration platforms). We hope that studies like ours presented here may help re-evaluate and re-design the algorithms and interfaces of online human communications and interactions for the betterment of our social network evolution.

This study is still limited in several aspects. First, we did not explore variations of the amplitude of stochastic fluctuations ($\epsilon$) or functional shapes of homophily and attention to novelty ($F_h$ and $F_a$). Second, transition planes were identified only by numerical simulations while analytical estimate of transition conditions is not accomplished yet. Third, we assumed that the behavioral parameter values would apply uniformly to all individuals in society with zero behavioral diversity. Fourth, the size of the simulated networks was relatively small (only up to 300 nodes). Future research directions are naturally to address each and all of these limitations in the current model. In particular, introducing individual behavioral diversity within a collective complex system is known to produce unexpected, nontrivial macroscopic outcomes \cite{sayama2020b,sayama2009}. Such behavioral heterogeneity should be represented in future models to gain more nuanced, more realistic collective outcomes. High Performance Computing frameworks for agent-based models \cite{paciorek2021} also may be used to increase the simulated network size and to expand parameter sweep ranges further. Finally, quantitative comparison and validation of model behaviors with actual social network evolution data will ultimately be needed. However, obtaining such empirical data of social network evolution has been extremely difficult, and this will remain one of the major challenges in adaptive social network modeling research.

\end{document}